%% file: top-2014-summary-CRM.tex
\documentclass[a4paper]{jpconf}
\usepackage{graphicx}
\usepackage{url,cite}

\def\gsim{\mathrel{\rlap{\lower4pt\hbox{\hskip1pt$\sim$}}
    \raise1pt\hbox{$>$}}}         
\def\lsim{\mathrel{\rlap{\lower4pt\hbox{\hskip1pt$\sim$}}
    \raise1pt\hbox{$<$}}}         

\begin{document}

\begin{flushright}
  Cavendish-HEP-14/15\\
  TTK-15-02
\end{flushright}
\vspace{-2.5cm}

\title{Summary of the Topical Workshop on Top Quark Differential Distributions 2014}

\author{Michal Czakon}
\address{Institut f\"ur Theoretische Teilchenphysik und Kosmologie,
  RWTH Aachen University, D-52056 Aachen, Germany}
\ead{mczakon@physik.rwth-aachen.de}
\author{Alexander Mitov}
\address{Cavendish Laboratory, University of Cambridge, Cambridge CB3 0HE, UK}
\ead{adm74@cam.ac.uk}
\author{Juan Rojo}
\address{Rudolf Peierls Centre for Theoretical Physics, 1 Keble Road, University of Oxford, OX1 3NP Oxford, UK}
\ead{juan.rojo@physics.ox.ac.uk}

\begin{abstract}
We summarise the {\it Topical Workshop on Top Quark Differential Distributions 2014}, 
which took place in Cannes immediately before the annual {\it Top2014} conference.
The
workshop was motivated by the availability of top 
quark differential distributions at NNLO and the forthcoming LHC 13 TeV data. 
The main goal of the workshop was to explore the impact of improved calculations of top quark production on precision 
LHC measurements, PDF determinations and searches for physics beyond the Standard Model,
as well as finding ways in which the high precision data from ATLAS, CMS and LHCb
can be used to further refine theoretical predictions for top production.
\end{abstract}

\section{Introduction}
The forthcoming availability of fully differential results for top quark production at NNLO,
together with the upcoming restart of the LHC at 13 TeV, prompted us to organise a workshop centred around theoretical aspects of precision top physics. The workshop was held 26--28 September 2014 in Cannes, i.e. it immediately preceded the annual Top2014 conference.

The workshop brought together a group of theory experts working on top quark physics and closely related subjects such as parton distribution functions, parton showers and soft gluon resummation. The presence of experts in BSM physics involving top quarks was essential, as was the participation of top quark experts from ATLAS, CMS and LHCb.

The main goal of the workshop was to explore the phenomenological implications of
the ongoing progress in precision calculations for top quark production, both
in terms of fully differential NNLO results and in terms of realistic description of
final states provided by NLO calculations matched to parton showers.

\begin{figure}[h]
\vspace{-4cm}
\begin{center}
\includegraphics[width=26pc]{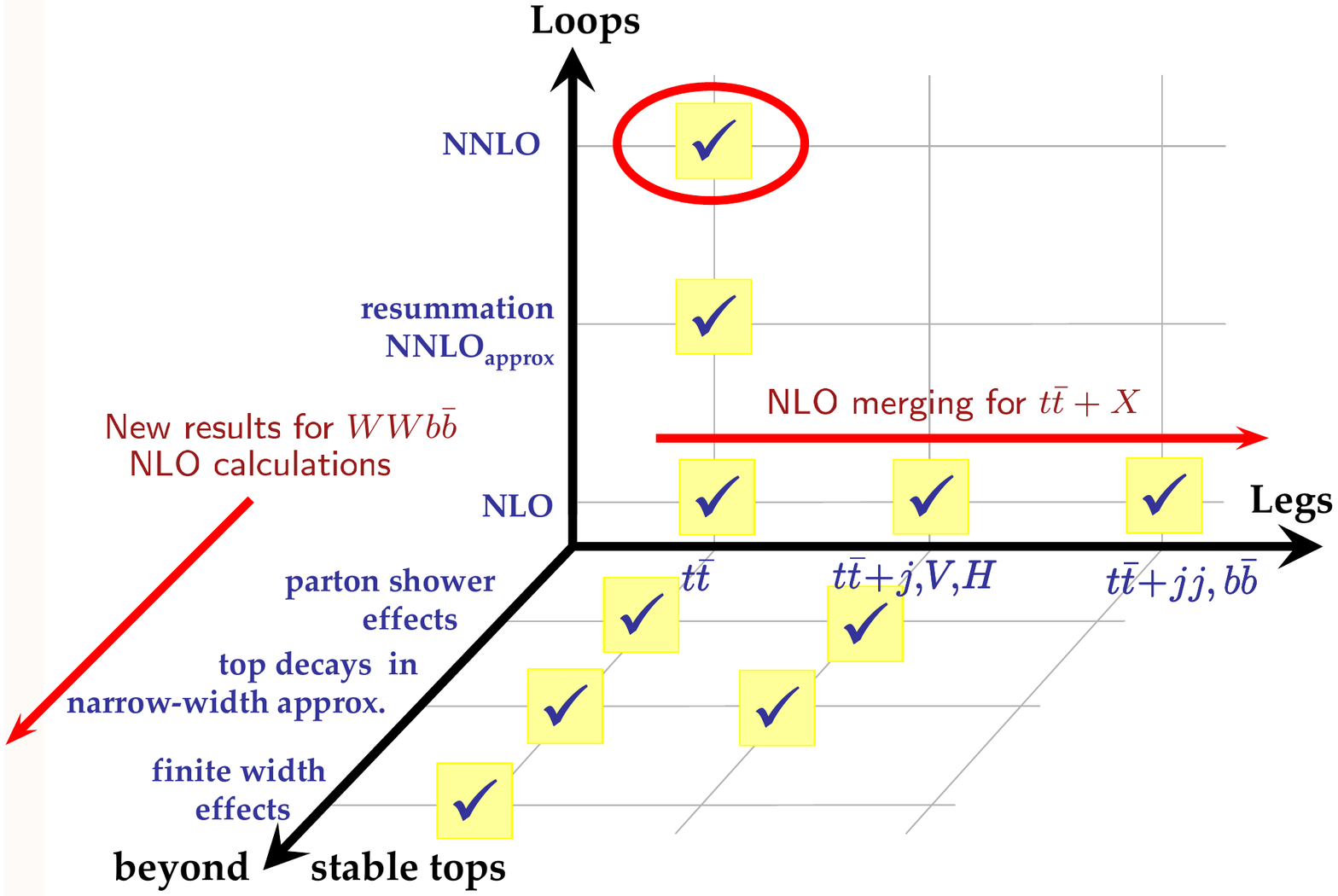}
\vspace{-3.7cm}
\caption{\label{scheme} \small Schematic illustration of recent theoretical
progress in theory calculations of top quark production in complementary
directions: NNLO calculations, merging of NLO+PS samples of different
multiplicities and results beyond stable tops (courtesy of M. Schultze).}
\end{center}
\end{figure}

The main questions that were discussed included: which top quark differential distributions
are theoretically and experimentally more interesting; what is the impact
of present and future top quark data on PDF fits; progress in realistic final states
with decaying top quarks; finite top quark width corrections and matching to parton
showers; the role of soft and collinear resummation; and
how precision calculations
in top quark physics can improve the reach of searches for BSM physics at the LHC.
Some of the most relevant recent theoretical developments are schematically
illustrated in Fig.~\ref{scheme}.
In addition, we convened a dedicated session on how precision theory developments can be used to 
improve ongoing and future measurements with top quarks at the LHC.

In the following, we present a concise summary of the discussions
that took place during the workshop. 
The complete agenda, with presentation slides, is available at the workshop webpage:
\begin{center}
  \url{http://indico.cern.ch/e/top-differential-distributions-2014}
\end{center}
Given the space limitation, we are unfortunately unable to cover in full detail everything that was discussed at the workshop.

\section{Top quark production and NNLO calculations}

\begin{figure}[h]
\vspace{-2.5cm}
\begin{center}
\includegraphics[width=18pc]{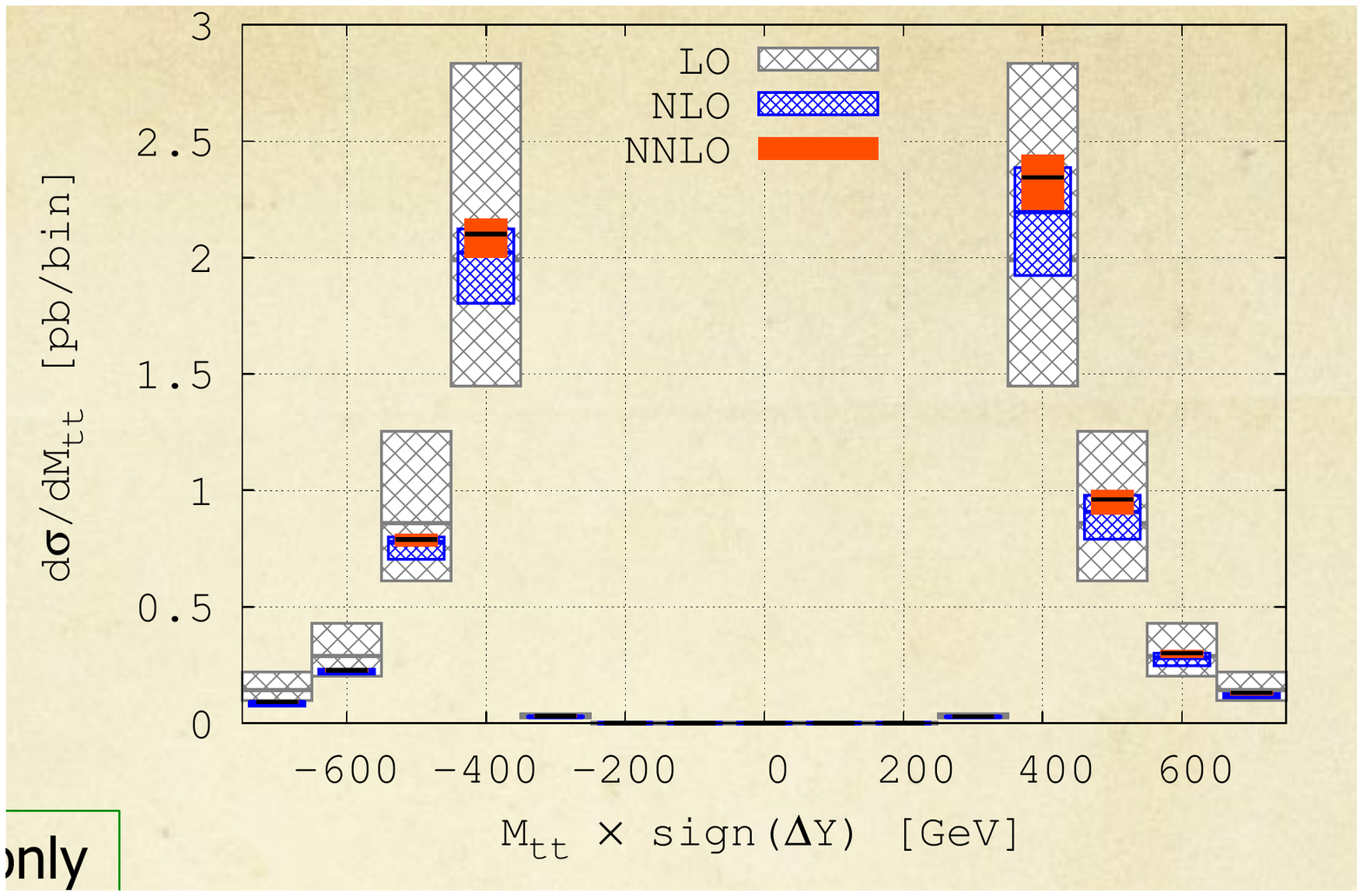}
\includegraphics[width=18pc]{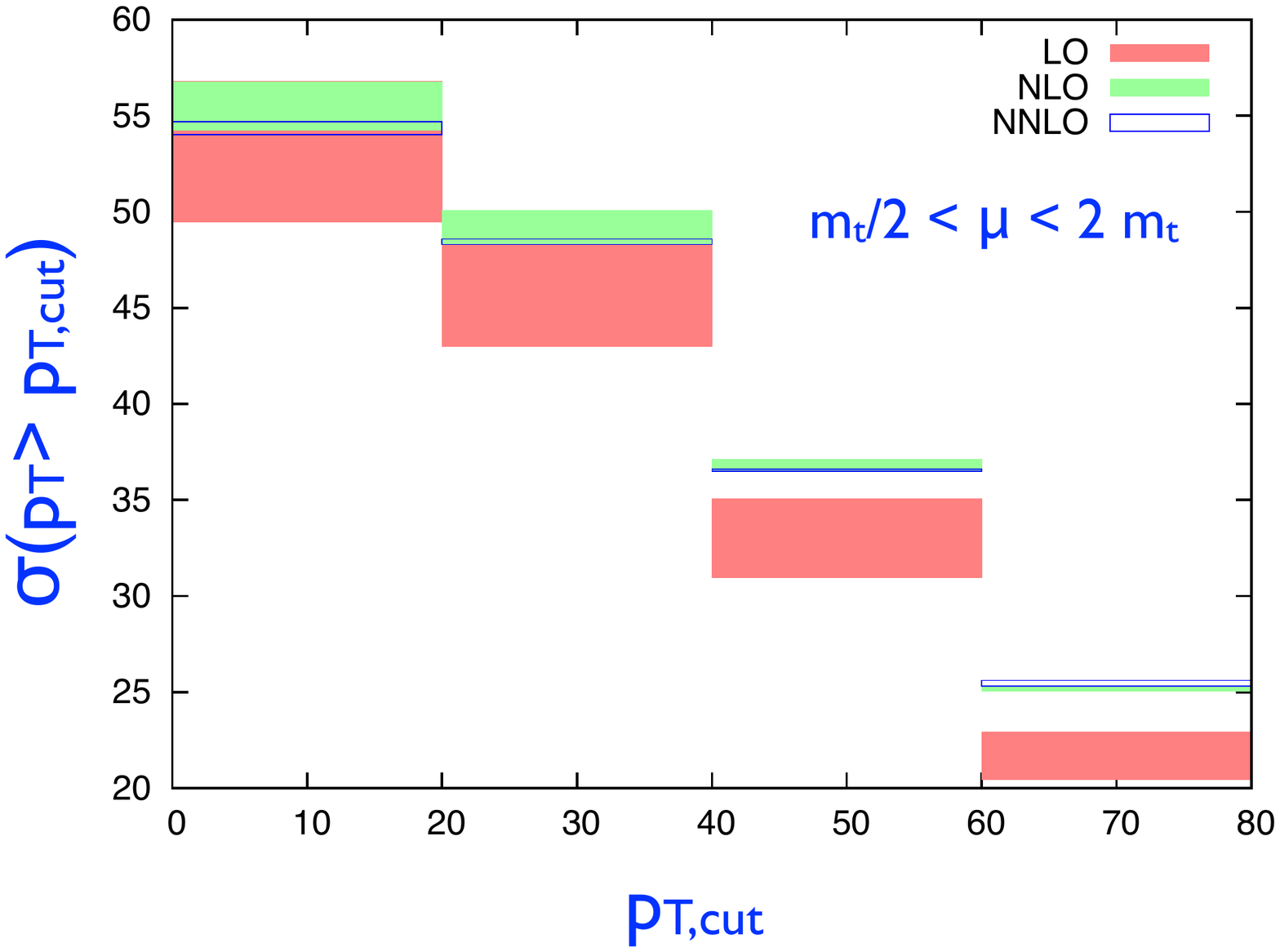}
\vspace{-3cm}
\caption{\label{nnlo} \small Left plot: preliminary results for
the invariant mass distribution $M_{tt}$ for top quark pair production at the
Tevatron at NNLO accuracy. Right plot: NNLO cross-section for the production of single top quarks
at the LHC 7 TeV, as a function of the cut in the $p_T$ of the top quark~\cite{Brucherseifer:2014ama}.}
\end{center}
\end{figure}

Recent progress in techniques for NNLO QCD calculations
\cite{GehrmannDeRidder:2005cm, Currie:2013vh, Catani:2007vq,
Czakon:2010td, Czakon:2011ve, Czakon:2014oma, Baernreuther:2013caa,
Henn:2013pwa} has led in the last few years to a 
dramatic increase in the availability of NNLO calculations for hadron collider processes with complex final states.
The NNLO result for the total top quark pair production cross-section has been
available for a while~\cite{Baernreuther:2012ws,Czakon:2012zr,Czakon:2012pz,Czakon:2013goa}. During
the workshop, preliminary results on the extension of this calculation to differential distributions
were presented.
These NNLO results are consistent with the NLO scale variation estimate,
and their inclusion leads to a significant
reduction of the theory uncertainties.
This is illustrated in Fig.~\ref{nnlo}, where we show the invariant mass
distribution $M_{tt}$ for top quark pair production at the
Tevatron at NNLO.
These results have also recently been used to compute the corrections to
the forward-backward asymmetry at the Tevatron in NNLO QCD \cite{Czakon:2014xsa}, showing
that perturbative corrections increase the absolute Standard Model (SM) value of the asymmetry by
about 2\%, thus improving the agreement of SM prediction with the measurements from CDF and D\O\ Collaborations.

Related techniques have also lead to the recent computation of the
NNLO corrections to single top production~\cite{Brucherseifer:2014ama}, which
was also discussed at the workshop.
As an illustrative result, in Fig.~\ref{nnlo} we show the cross-section
as a function of the cut in the $p_T$ of the top quark, for the LHC 7 TeV.
For this observable the NNLO corrections also lead to a substantial improvement in the perturbative
expansion.
Another central process of the LHC program is dijet production, due to its
relevance for precision Standard Model measurements, PDF determinations
and new physics searches.
Recent results towards the full NNLO calculation~\cite{Ridder:2013mf}
were presented in the workshop.
Several other important LHC processes have also become available at NNLO, see
for instance~\cite{Boughezal:2013uia,Gehrmann:2014fva,Cascioli:2014yka}. 
More processes/observables will be computed in the near future, underscoring the trend towards NNLO QCD becoming
the standard for precision phenomenology at the LHC.
However, work is still required to be able to use these calculations with
realistic final states, as we report below.

\section{Top pair production with realistic final states}

\begin{figure}[h]
   \vspace{-2cm}
 \begin{center}
  \includegraphics[width=17pc]{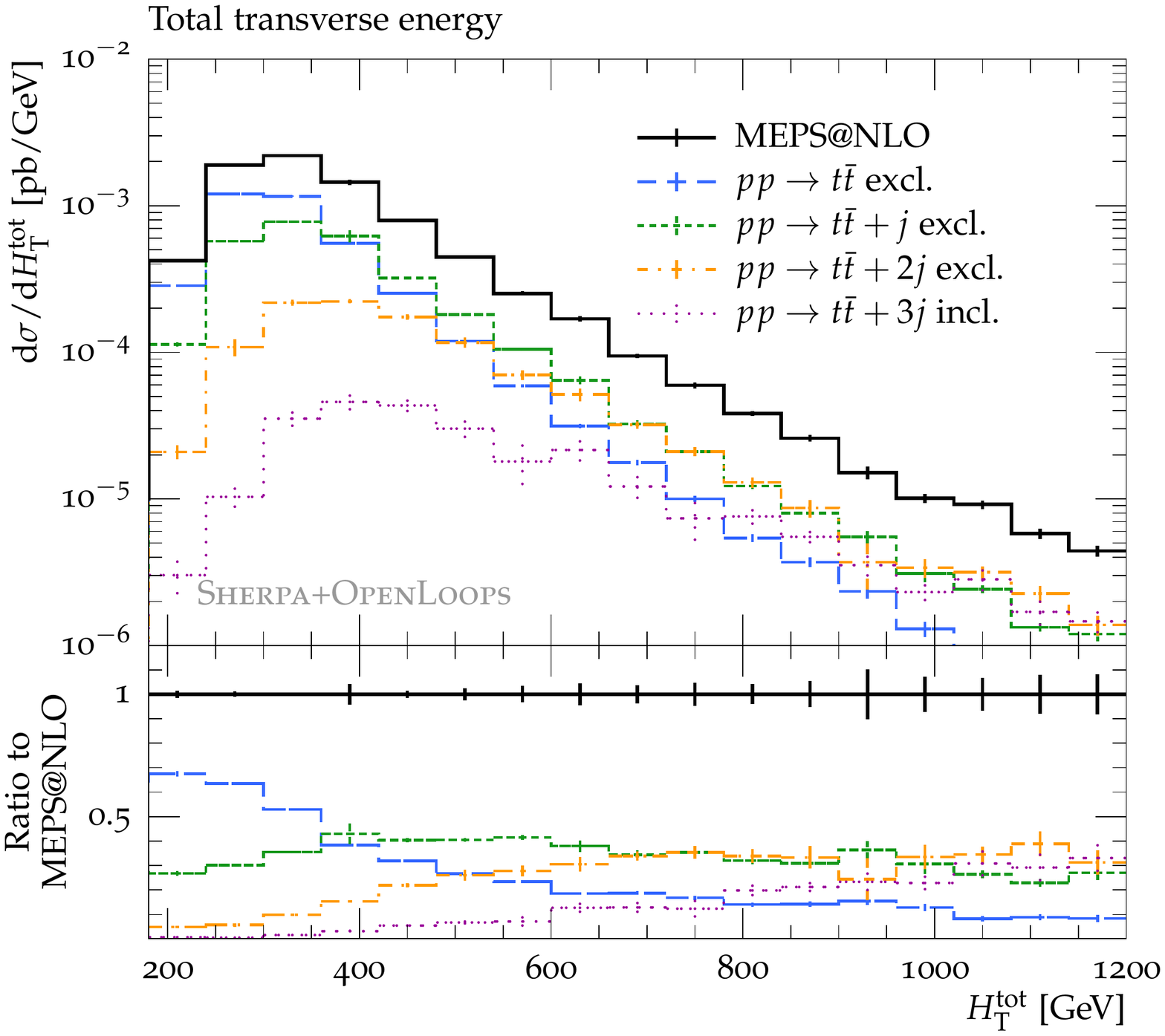}
  \includegraphics[width=17pc]{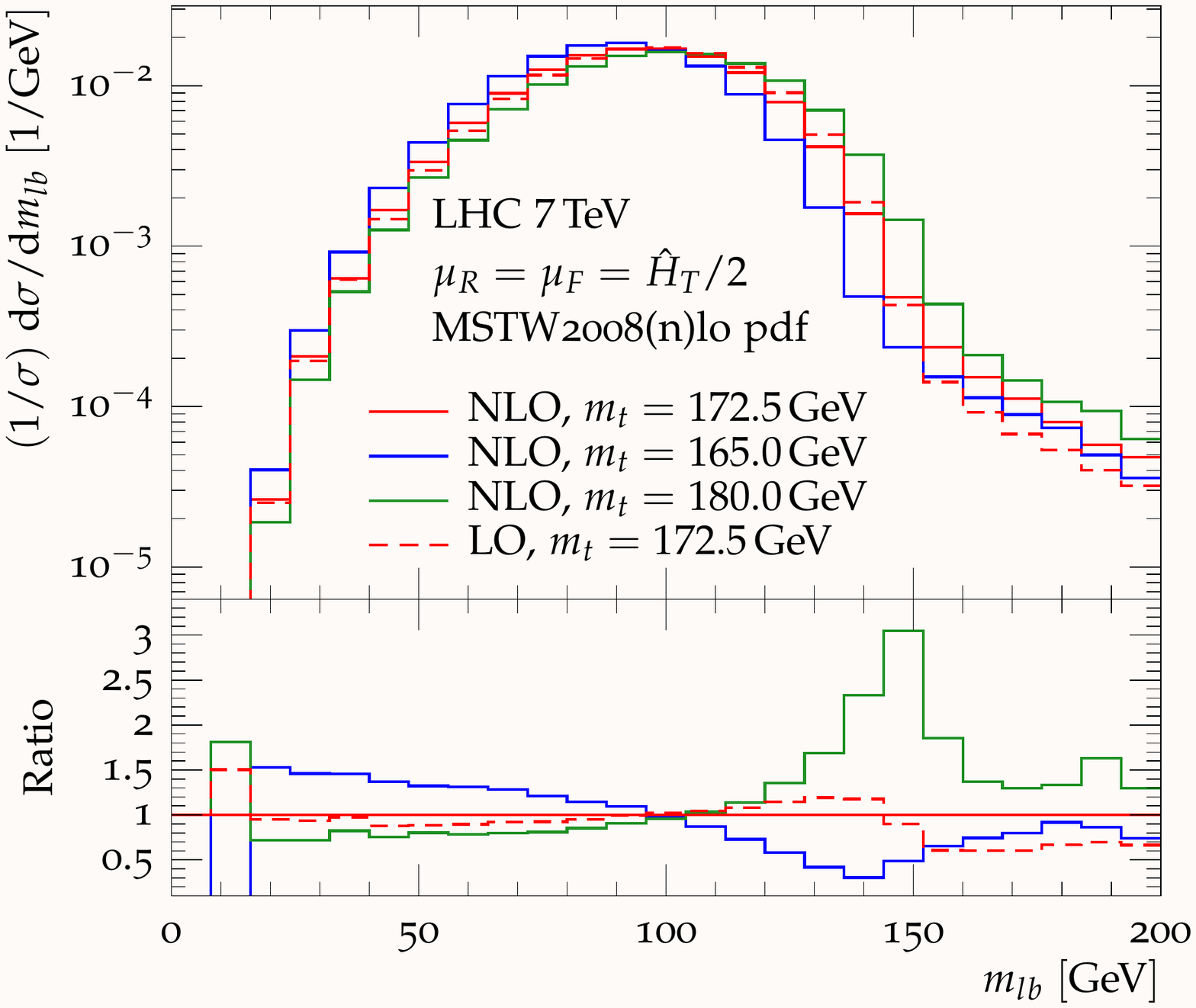}
  \vspace{-2.5cm}
  \caption{\label{merging} \small Left plot: $H_T$ distribution in $t\bar{t}$+jets events at the LHC 7 TeV
    in the MEPS@NLO approach, compared to samples with exclusive jet multiplicities.
    Right plot: the $m_{lb}$ distribution computed at NLO at the
    LHC 7 TeV, for different values of the top quark mass.
}
\end{center}
\end{figure}

The current state-of-the-art simulations of top quark
production and decay utilize merged NLO calculations matched to parton showers.
Various  proposals for NLO merging have been introduced recently,
including the FxFx merging~\cite{Frederix:2012ps}, the UNLOPS procedure~\cite{Lonnblad:2012ix}
and the MEPS@NLO~\cite{Hoeche:2012yf} method among others.
The implications of some of these updated calculations for top quark pair production were discussed at the workshop.
As a representative result, in Fig.~\ref{merging} we show the $H_T$ distribution in $t\bar{t}$+jets events at LHC 7 TeV
    within the MEPS@NLO approach, compared to predictions based on
    samples with exclusive jet multiplicities.
    Clearly, the individual multiplicities contribute differently
    depending on the value of $H_T$, while the NLO merged sample
    could be applied to the full phase space.
    There has also been important progress in the matching of NNLO calculations
    to parton showers~\cite{Hoche:2014dla,Karlberg:2014qua}, though
    still quite some work is needed to be able
    to apply these methods to top quark production.
    Some preliminary results for matching to the Nagy-Soper parton
    shower with quantum interference \cite{Nagy:2007ty} at NLO have been
    presented \cite{CHKW}. Though they are still restricted to on-shell
    top-quarks, work in the direction of realistic final states is
    under way.

Another important application of the recent progress in top physics calculations
is to precision top quark mass determination.
For instance, the top quark mass $m_t$ can be extracted from template fits to the
$m_{lb}$ invariant mass distribution with good experimental precision.
However, unless the NLO corrections to the $pp\to WbWb$ process are
accounted for, theoretical uncertainties due to missing higher orders will
dominate the total $m_t$ uncertainty.
In Fig.~\ref{merging} we show  the $m_{lb}$ distribution computed at NLO at the
    LHC 7 TeV, for different values of the top quark mass~\cite{Heinrich:2013qaa}.
    Similar conclusions, of course, also applies to many other
    differential distributions whose accurate prediction is an important ingredient in new physics searches.

\section{Experimental issues}

One of the workshop's main objectives was to initiate a discussion about how ongoing theoretical progress and current and future experimental measurements can improve each other.
Some of the discussed issues were triggers, pile-up subtraction,
theory uncertainties that affect the selection efficiency, improvements
in Monte Carlo simulations, definitions of physical observables
and top quark reconstruction.

In this context, one of the most important topics for discussion was how
the availability of NNLO top quark differential distributions, as well as improvements
in the description of realistic final states at NLO, can help in reducing various extrapolation
errors.
Since state-of-the-art MC simulations are now based on NLO calculations that are merged for different topologies and matched to parton showers, it will be essential for precision top physics at LHC Run II to adopt these tools as standard. 
The availability of differential predictions, either NNLO with stable tops
or NLO+PS with realistic final states, represents a strong motivation for all LHC measurements
to be provided directly in the fiducial region, and that comparisons with theory
are performed at this level.
Such a ``meeting point" between theory and experiment avoids inconsistencies in various comparisons or, equivalently, the unnecessary increase of theoretical uncertainties in the extrapolation to the full phase space. 
Of course inclusive measurements are also important in many cases, for example
in comparisons with other experiments, but the original information in the fiducial
region should also be available.
In this respect, an important improvement in realistic analyses would be the extension
of the NNLO calculation to the case of unstable tops.

An important ingredient in top production predictions made with MC event generators is the tune for the semi-hard and soft physics.
Such tunes are typically performed with LO Monte Carlos (see for example the
recent {\sc\small Monash} 2013 Tune~\cite{Skands:2014pea} of {\sc\small Pythia8}~\cite{Sjostrand:2007gs})
and then applied to NLO+PS generators.
Given the importance of non-perturbative and semi-hard physics in various top quark
measurements, it would be
of utmost importance to produce dedicated new tunes for NLO
generators that are able to describe simultaneously the hard, semi-hard and soft dynamics.
Work along these lines is ongoing within the ATLAS and CMS collaborations, with the aim of
obtaining dedicated NLO tunes that can then be applied to top physics at Run II.

Also discussed at the workshop was the possibility
to present top measurements in terms of ratios of various cross-sections,
in order to partially cancel some of the leading experimental
and theoretical uncertainties.
For instance, in early Run II data the LHC luminosity uncertainty could be
substantial, and measurements of ratios like $\sigma(t\bar{t})/\sigma(Z)$
should
allow to perform precision top quark physics already
from the first months of data taking.
Related proposals include ratios such as $\sigma(t\bar{t}b\bar{b})/\sigma(t\bar{t}jj)$,
which also provide stringent tests of MC event generators.

In addition, the ratios of top quark cross-sections between 13 TeV and
8 TeV provide a unique opportunity to constrain the gluon PDF with greatly reduced theory
uncertainties from scale and $m_t$.
The advantages of presenting the measurements of top quark differential distributions either with absolute normalization or normalized to the fiducial cross-section were also discussed.
The consensus in the community is that the measurements should be presented both ways, i.e.
if normalized measurements are published, the normalization factor should be provided as well.
We recall that while for many analyses (for example searches) only an accurate
measurement of the shape of the distribution is required, for others (in particular
PDF analyses) the overall normalization provides precious additional information.

An important topic of the discussion was how to optimize the use of theoretical
calculations when kinematical distributions are used to extract SM parameters
such as $\alpha_S(M_Z)$ and $m_{t}$.
For instance, CMS has extracted $\alpha_S(M_Z)$ from the inclusive $t\bar{t}$ cross-sections
using the inclusive NNLO calculation~\cite{Chatrchyan:2013haa} and it would be interesting to repeat the extraction
from differential distributions.
Concerning $m_t$ extractions, it became clear that it is essential to quantify the
theory uncertainties from template fits of kinematical distributions;
in particular NLO QCD should be the baseline for the computation
of these templates (since LO calculations have large associated scale uncertainties).
The use of double differential measurements could be beneficial here, provided one is not limited by statistics.

Finally, we discussed the fact that many searches that utilise measurements
with top quarks in the final state are never unfolded and recast
in terms of differential SM measurements.
Translating searches into SM measurements could be very beneficial since, first, these allow new precision SM studies in extreme kinematical regions, and second, existing searches could easily be re-applied to different BSM scenarios.

\section{Top quark data and PDF fits}

At the LHC, top quark pairs are produced predominantly in the gluon-gluon
initial state.
Therefore, the recent improvements in the precision of both experimental
measurements and theoretical calculations for top pair production
strongly suggest that top data should provide useful constraints
on the poorly-known large-$x$ gluon PDF, fully complementary to those
obtained from other processes like jet production~\cite{Rojo:2014kta} or photon production~\cite{d'Enterria:2012yj,Carminati:2012mm}.
Several studies have demonstrated that already at the level of inclusive cross-sections,
available top quark data from ATLAS and CMS at 7 TeV and 8 TeV can provide important
information on the gluon for $x \gsim 0.1$~\cite{Alekhin:2013nda,Czakon:2013tha,Beneke:2012wb}.
In addition, the feasibility of top quark production in the forward region by LHCb
and the possible constraints in PDFs that such data would provide
has also been quantified~\cite{Gauld:2013aja}.
Total cross-sections for $t\bar{t}$ production are already included in the recent
NNPDF3.0~\cite{Ball:2014uwa} and MMHT14~\cite{Harland-Lang:2014zoa} global analysis,
and are also available
in the {\sc\small HERAfitter} open-source QCD fit framework~\cite{Alekhin:2014irh}.

\begin{figure}[h]
   \vspace{-2.9cm}
 \begin{center}
  \includegraphics[width=17pc]{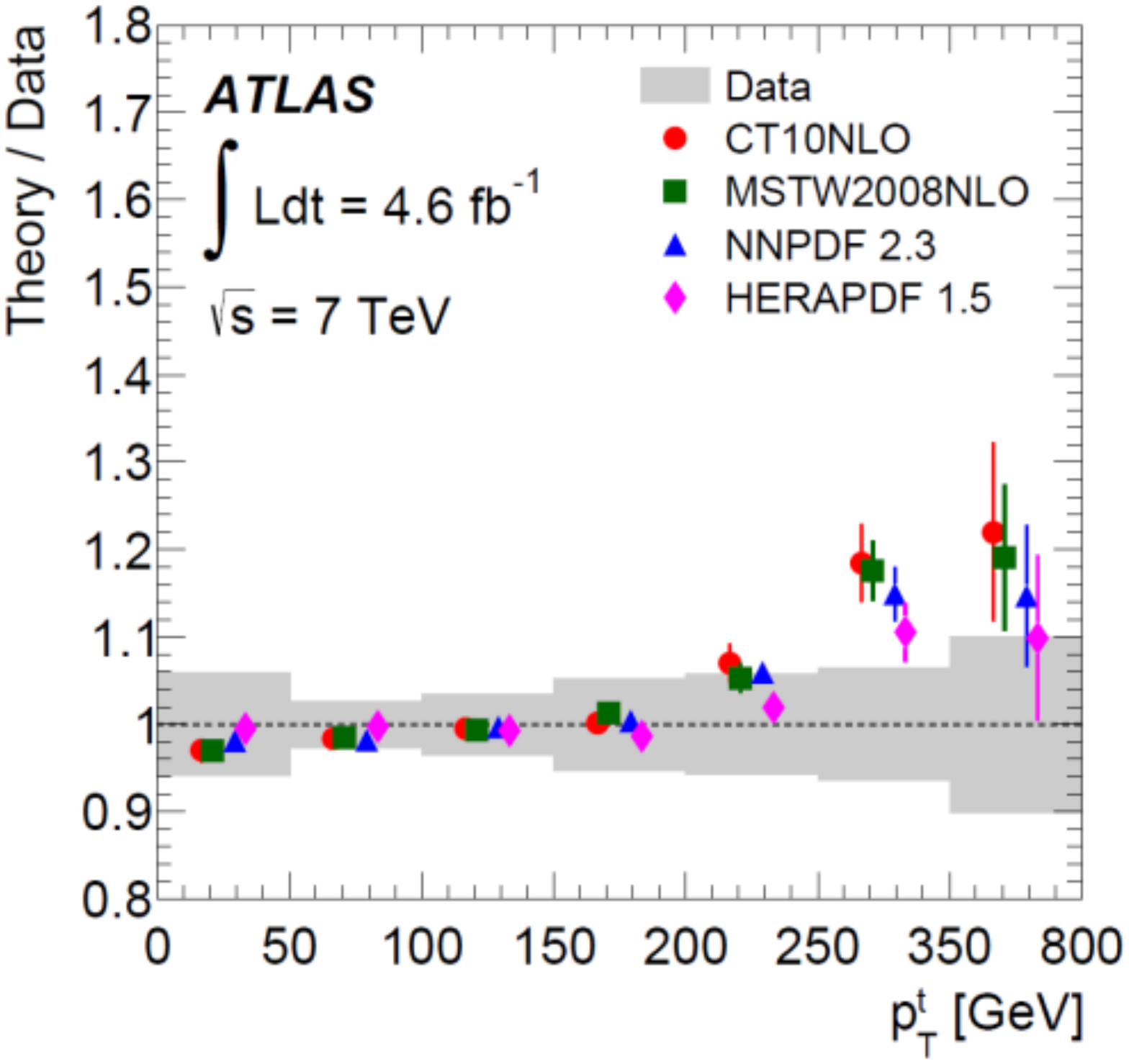}
  \includegraphics[width=19pc]{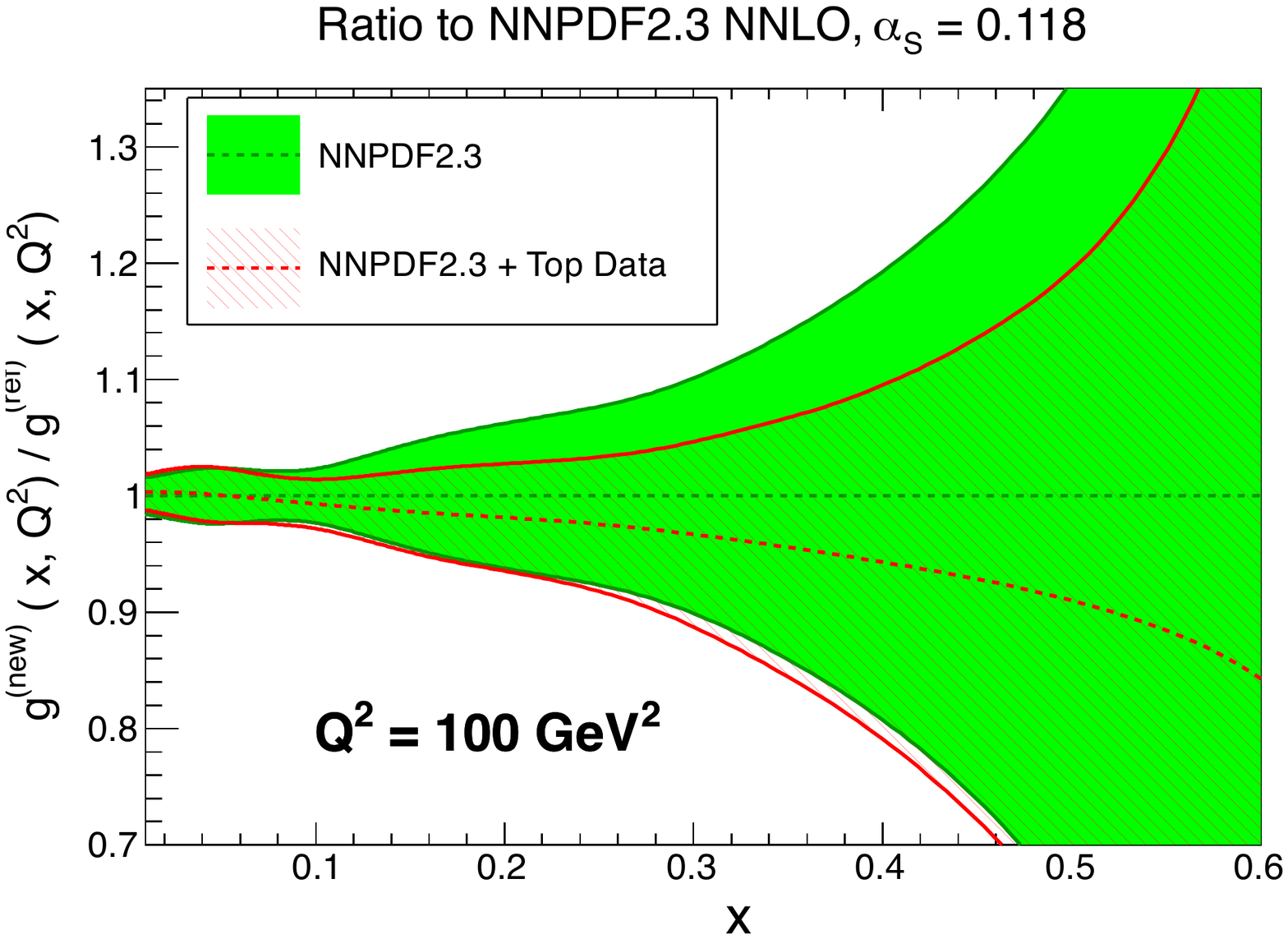}
  \vspace{-2.5cm}
  \caption{\label{fits} \small Left plot: the 7 TeV
    ATLAS measurements of the $p_T$ of top quarks,
    compared to various PDF sets.
    Right plot: the reduction of the gluon large-$x$
    PDF uncertainties when inclusive top quark cross-sections
    are added to the NNPDF2.3 NNLO fit~\cite{Ball:2012cx}.
}
\end{center}
\end{figure}

The challenge now is to include the differential distributions of top pairs
into global PDF analysis, using both the NNLO results and the recent availability
of experimental measurements from ATLAS and
CMS~\cite{Chatrchyan:2012saa,Aad:2014zka,Aad:2012hg}.
In Fig.~\ref{fits}
we show the recent ATLAS 7 TeV differential measurements~\cite{Aad:2014zka}
compared with different PDF sets: including these data into PDF fit will provide
a handle on the large-$x$ gluon.
In exploiting these measurements, it will be necessary to exploit the full
power of recent theory calculation for these distributions, by combining fast
interfaces to NLO calculations like {\sc\small aMCfast}~\cite{amcfast},
where the {\sc\small APPLgrid}~\cite{Carli:2010rw}
framework is used to precompute {\sc\small  Madgraph5\_aMC@NLO}~\cite{Alwall:2014hca}
cross-sections, with the exact NNLO results.
The same fast grid techniques could also be used to interface directly
the very CPU-intensive NNLO calculations into PDF fits.

Another interesting observable that was discussed is the ratio of cross-sections between 13
TeV and 8 TeV~\cite{Mangano:2012mh},
where many experimental and theory systematics cancels, providing a clean
handle on PDFs.
In the specific case of top quark production, the dependence on $m_t$
and on the scales is largely canceled
in such a ratio, and theory uncertainties are mostly driven by differences
in the gluon PDF.
There are plans to perform these measurements by both ATLAS and CMS.

\section{Approximate calculations for top pair production}

In addition to exact NLO and NNLO calculations (both fixed order
and matched to parton showers), we also discussed recent progress in
approximate calculations in top quark production.
In the case of the total cross-section, it has been recently proposed that
it is possible to compute a robust estimate of yet unknown higher orders
by using known results and exploiting the analytic properties of the partonic
cross-sections in Mellin space.
This technique has been successfully applied to Higgs production in gluon
fusion~\cite{Ball:2013bra}, validated by available NNLO and partial N3LO results,
and during the workshop we discussed its extension to an approximate
N3LO $\sigma_{t\bar{t}}$ calculation.
An interesting related issue with approximate N3LO calculations is whether NNLO PDFs are sufficient, or if one really needs N3LO PDFs.
This problem has been addressed in Ref.~\cite{Forte:2013mda}, finding that, interestingly, N3LO
are not required for Higgs production, but that they are needed for
$t\bar{t}$ production.
The reason for this is the fact that in top production a larger
value of Bjorken-$x$ is probed that in Higgs production.

For differential distributions, approximate higher-order results can be obtained using
techniques that stem from the resummation to all orders of
terms enhanced in the soft and collinear limits.
For instance, we discussed recent studies~\cite{Broggio:2014yca},
where the renormalization group equations are used to
derive
approximate NNLO top quark differential distributions
including semi-leptonic top quark decays computed in the narrow-width approximation.
Related earlier studies include~\cite{Ahrens:2010zv}.
Previously, various approximations for the NNLO cross-section 
were also available~\cite{Cacciari:2011hy,Beneke:2012wb}.

\section{BSM physics searches with top quarks}

Top quarks are a crucial ingredient in essentially all scenarios
for physics beyond the Standard Model.
Their large Yukawa coupling suggest that they could play a major role
in understanding the origin of the electroweak symmetry breaking
mechanisms.
In addition, the top quark contribution to quantum corrections to the Higgs boson mass
is at the heart of the hierarchy problem, and naturalness-based solutions to this
problem typically require the presence of top partners.
It is therefore clear that searches for New Physics that involve top quarks
in the final state are very important at the LHC.

\begin{figure}[h]
   \vspace{-2cm}
 \begin{center}
  \includegraphics[width=17pc]{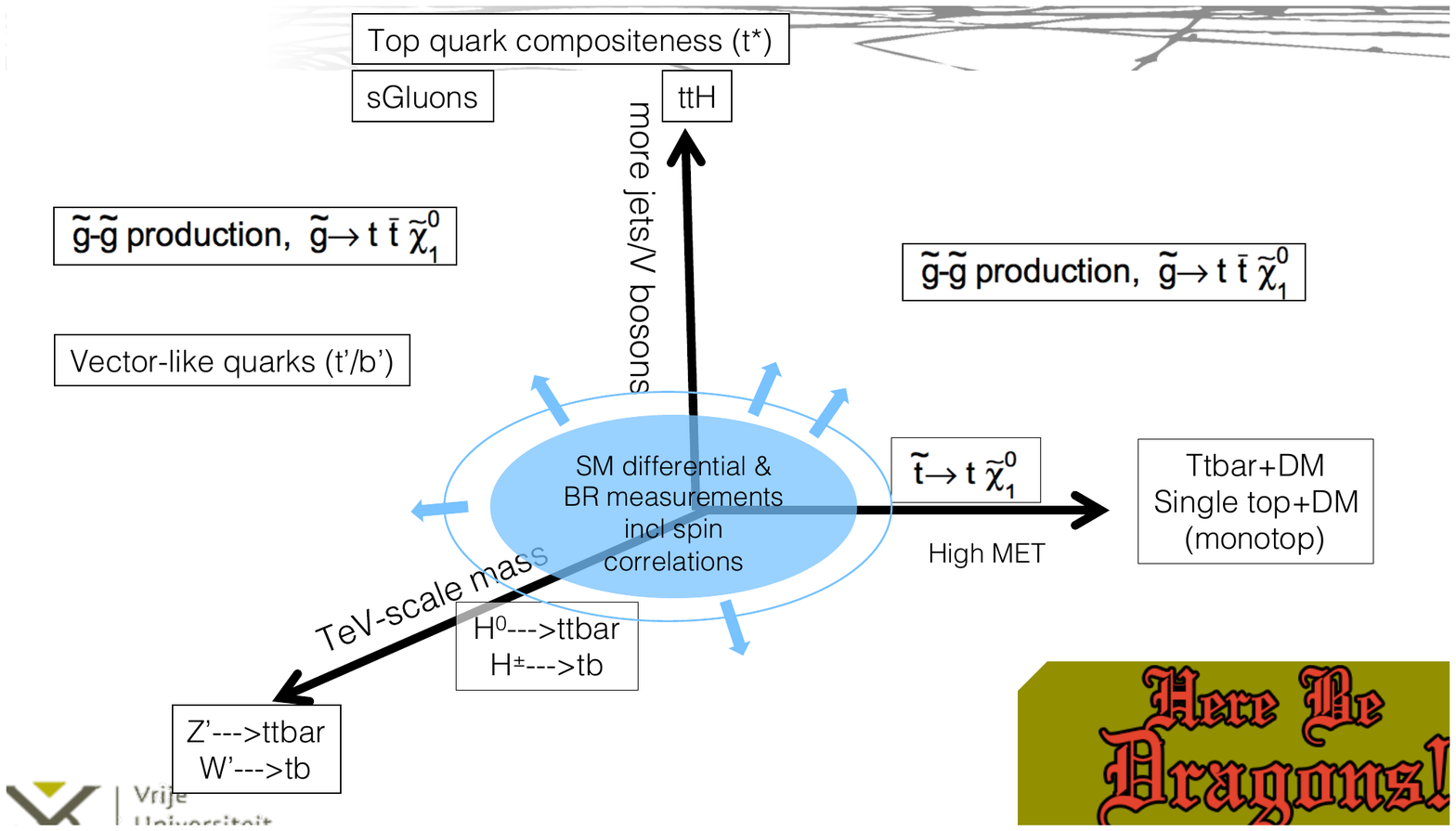}
  \includegraphics[width=17pc]{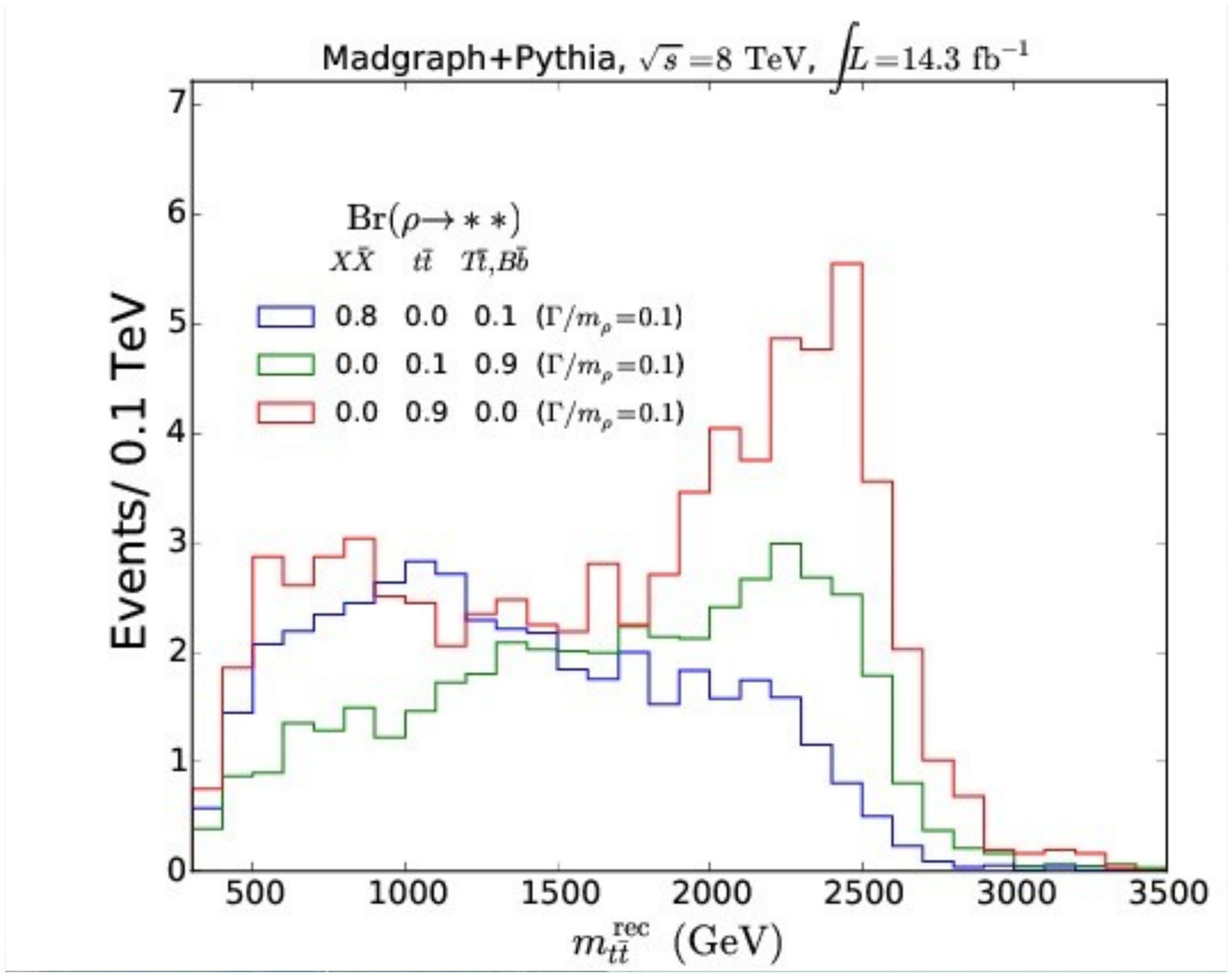}
  \vspace{-2.5cm}
  \caption{\label{bsm} \small Left plot: schematic illustration
    of the different extreme kinematical regions probed by
    BSM searches at the LHC with top quarks in the final state (courtesy of F. Blekman).
    Right plot: the reconstructed mass of a top quark pair $m_{tt}^{\rm reco}$
    in BSM scenarios where a heavy resonance $\rho$ decays into a $t\bar{t}$ pair
    (red histogram) or to other intermediate resonances which in turn decay to
    top quarks (blue and green histograms), from~\cite{Chala:2014mmam}.
}
\end{center}
\end{figure}

In these searches, SM top quark production is typically the dominant background,
and therefore the recent developments in precision calculations in top physics
should certainly improve the reach of BSM searchers.
For instance, the NNLO calculation of the total cross-section has been
used in~\cite{Czakon:2014fka,Aad:2014kva} to
improve the bounds on light stop quarks.
Another example is provided by the fact that PDF uncertainties are one of the
dominant modeling systematics in many searches, specially those that
involve invariant masses at the TeV scale, and reducing these PDF errors, with
top quark data in particular,
would certainly improve the reach of these searches for New Physics~\cite{Czakon:2013tha}.
Another important aspect that was emphasized during the workshop was that BSM searches
typically probe top quark production in extreme kinematical regions, as
illustrated schematically in Fig.~\ref{bsm}.
In particular, BSM searches require good understanding of top quark production
in association with many jets or vector bosons, top quark pairs
with TeV invariant masses and $t\bar{t}$ and single top production in association
with substantial missing $E_T$.

Another topic that was discussed is that recent LHC results strongly suggest that
one should adopt new search strategies to look for New Physics.
For instance, many searches look for heavy resonances coupled to top quarks
by reconstructing the invariant mass of the $t\bar{t}$ pair and trying to
identify a resonance on top of the SM background.
However, in more realistic BSM scenarios this heavy resonance will instead decay
to other BSM states which in turn decay to top quarks, and the peak in $t\bar{t}$
will disappear, see Fig.~\ref{bsm} for an illustration taken from~\cite{Chala:2014mmam}.
This example shows that a cross-talk between BSM theorists, theorists
involved in precision SM calculations and experimentalists is essential in order to
to maximize the scientific output of the LHC data analyses.

\section*{Acknowledgments}
We warmly acknowledge all workshop participants for their
enthusiastic contributions and for the many lively discussions.
We are especially grateful to Frederic Deliot and Roberto Chierici for
their help with the organization of the workshop. 
The work of M.~C. was supported by the German Research Foundation (DFG) via the Sonderforschungsbereich/Transregio SFB/TR-9 ``Computational Particle Physics". The work of A.~M. is supported by the UK Science and Technology Facilities Council [grants ST/L002760/1 and ST/K004883/1].
The work of J.~R. was supported by an STFC Rutherford Fellowship ST/K005227/1 and by the European Research Council Starting Grant "PDF4BSM".

\section*{References}
\input{top-2014-summary-CRM.bbl}

\end{document}

%% file: top-2014-summary-CRM.bbl
\providecommand{\newblock}{}